\documentclass[conference]{IEEEtran}
\IEEEoverridecommandlockouts
\usepackage{cite}
\usepackage{amsmath,amssymb,amsfonts}
\usepackage{algorithmic}
\usepackage{graphicx}
\usepackage{textcomp}
\usepackage{xcolor}

\usepackage{stfloats}
\usepackage{verbatim}
\usepackage{graphicx}
\usepackage{cite}
\usepackage{makecell}
\usepackage{threeparttable}
\usepackage{array}
\usepackage{booktabs}
\usepackage{multirow}
\usepackage{stfloats}
\usepackage{bm}
\usepackage{booktabs}

\def\BibTeX{{\rm B\kern-.05em{\sc i\kern-.025em b}\kern-.08em
    T\kern-.1667em\lower.7ex\hbox{E}\kern-.125emX}}
\begin{document}

\title{A 0.96pJ/SOP, 30.23K-neuron/mm$^2$ Heterogeneous Neuromorphic Chip With Fullerene-like Interconnection Topology for Edge-AI Computing
}

\author{P. J. Zhou, Q. Yu, M. Chen, Y. C. Wang, L. W. Meng, Y. Zuo, N. Ning, Y. Liu, S. G. Hu, G. C. Qiao$^*$

\thanks{P. J. Zhou, Q. Yu, M. Chen, Y. C. Wang, L. W. Meng, Y. Zuo, N. Ning, Y. Liu, S. G. Hu and G. C. Qiao are with the State Key Laboratory of Electronic Thin Films and Integrated Devices, University of Electronic Science and Technology of China, Chengdu 610054, P. R. China.}
\thanks{$^*$Corresponding authors: G. C. Qiao (e-mail: gcqiao@uestc.edu.cn)}
}



\maketitle

\begin{abstract}
    Edge-AI computing requires high energy efficiency, low power consumption, and relatively high flexibility and compact area, challenging the AI-chip design. This work presents a 0.96 pJ/SOP heterogeneous neuromorphic system-on-chip (SoC) with fullerene-like interconnection topology for edge-AI computing. The neuromorphic core integrates different technologies to augment computing energy efficiency, including sparse computing, partial membrane potential updates, and non-uniform weight quantization. Multiple neuromorphic cores and multi-mode routers form a fullerene-like network-on-chip (NoC). The average degree of communication nodes exceeds traditional topologies by 32\%, with a minimal degree variance of 0.93, allowing advanced decentralized on-chip communication. Additionally, the NoC can be scaled up through extended off-chip high-level router nodes. A RISC-V CPU and a neuromorphic processor are tightly coupled and fabricated within a 5.42 mm$^2$ die area under 55 nm CMOS technology. The chip has a low power density of 0.52 mW/mm$^2$, reducing 67.5\% compared to related works, and achieves a high neuron density of 30.23 K/mm$^2$. Eventually, the chip is demonstrated to be effective on different datasets and achieves 0.96 pJ/SOP energy efficiency. 
\end{abstract}


\section{Introduction}
Neuromorphic computing using spiking neural networks (SNNs) has proven to be efficient, low power, and hardware-friendly, which has the potential to be deployed at the edge platform for edge-AI computing\cite{b1,b2,b3}. Although several dedicated neuromorphic chips have been proposed, leading to impressive achievements, deploying them on edge devices remains challenging due to their monofunctional nature\cite{b4,b5,b6,b7,b8}. 

For multifunctionality, heterogeneous neuromorphic chips with multiple dedicated computing cores have been developed\cite{b9,b10,b11}. To enhance the flexibility of neuromorphic chips in edge devices, it is imperative to further realize the heterogeneous integration of dedicated and general-purpose computing units. However, a challenge exists in establishing efficient coupling between heterogeneous cores. Furthermore, previous works still suffer from inefficiencies in on-chip communication and data reuse as well as low neuron integration, which are challenging to realize edge-AI computing within constrained power and area.

To address these challenges, this work reports a low-power heterogenous neuromorphic system-on-chip (SoC) with a RISC-V CPU and a neuromorphic processor for edge-AI computing, which is deeply optimized from core-, network-on-chip- and system-levels. Sparse spike zero-skip, partial membrane potential (MP) update, weight non-uniform quantization, partial synapse parallel process, and pipeline design are adopted to improve computing efficiency while reducing power consumption and hardware cost. A fullerene-like NoC with high connectivity and low latency is designed for high-performance on-chip communication. An extended neuromorphic unit (ENU) is adopted to couple the RISC-V CPU and neuromorphic processor. The key characteristics of the chip are listed as follows:
\begin{itemize}
    \item The neuromorphic core achieves a synapse energy efficiency of $\leq$1.196 pJ/SOP and a computing efficiency of $\geq$0.426 GSOP/s when the spike sparsity exceeds 40\%. Its energy efficiency improves $\times$2.69 than the traditional scheme.
    \item The fullerene-like NoC achieves a low average latency of 3.16 hops, 39.9\% less than other NoCs. The average degree of communication nodes exceeds traditional topologies by 32\%, with a minimal degree variance of 0.93, allowing advanced decentralized on-chip communication. Additionally, a multi-mode router realizes a maximum bandwidth of 0.4 spike/cycle with a sub-0.1 pJ/hop transmission energy efficiency.
    \item The RISC-V works at a low average power of 0.434 mW, 43\% decrease from the baseline. Meanwhile, the ENU tightly couples the RISC-V CPU and neuromorphic processor through extended instructions, enabling efficient interaction among heterogeneous cores.
    \item The chip is fabricated on 3.41 mm$^2$ (without pad) under 55 nm CMOS technology, which is demonstrated effective on different datasets and achieves the best 0.96 pJ/SOP on the NMNIST dataset. It has 2.8 mW low power and 0.52 mW/mm$^2$ power density, better than previous works. 
\end{itemize}

\section{Heterogeneous Neuromorphic Chip Design}
\subsection{Neuromorphic Core Design}
\begin{figure}[!t]
    \centerline{\includegraphics[width=\columnwidth]{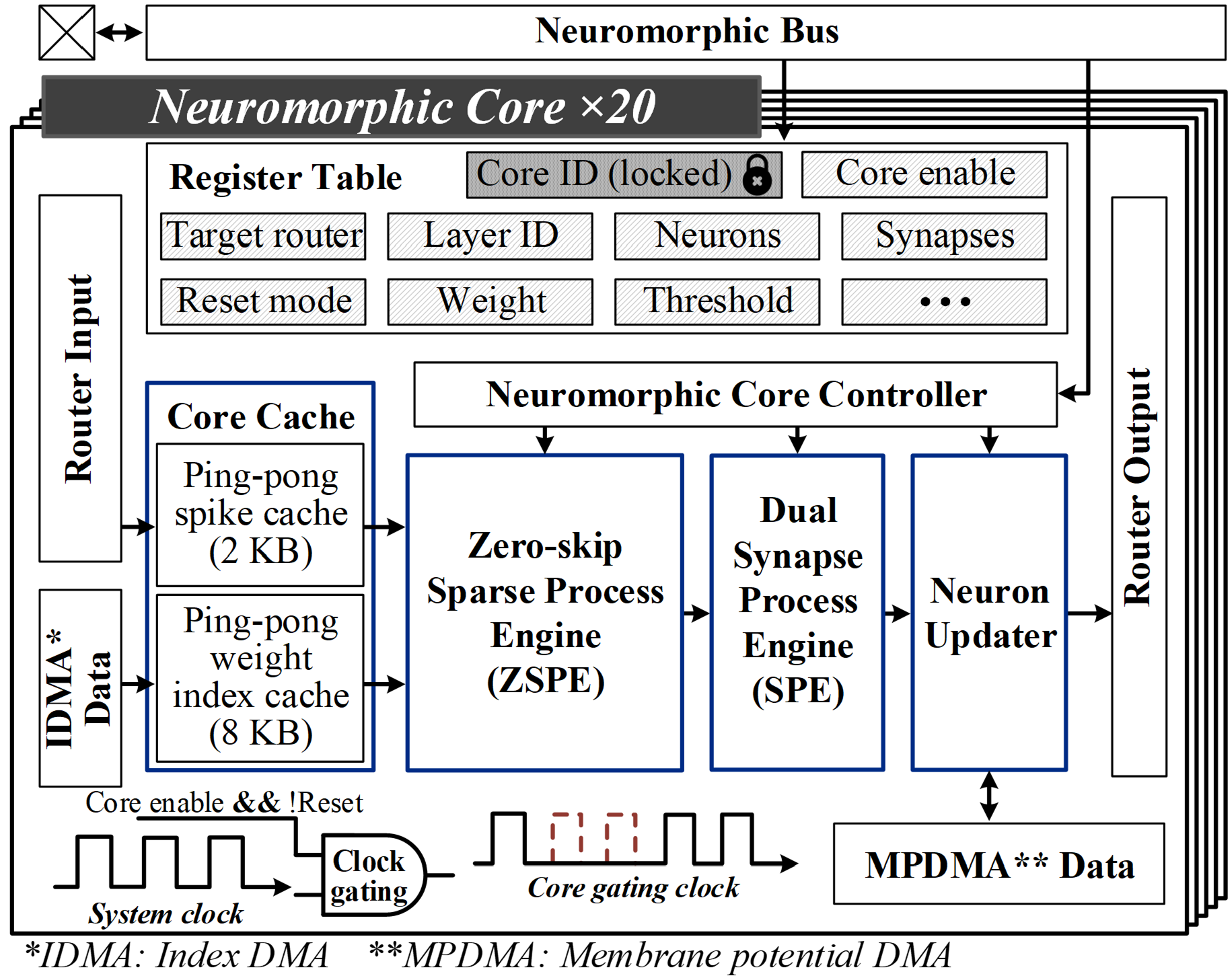}}
    \caption{Scheme of the neuromorphic core.}
    \label{fig1}
\end{figure}
\begin{figure}[!t]
    \centerline{\includegraphics[width=\columnwidth]{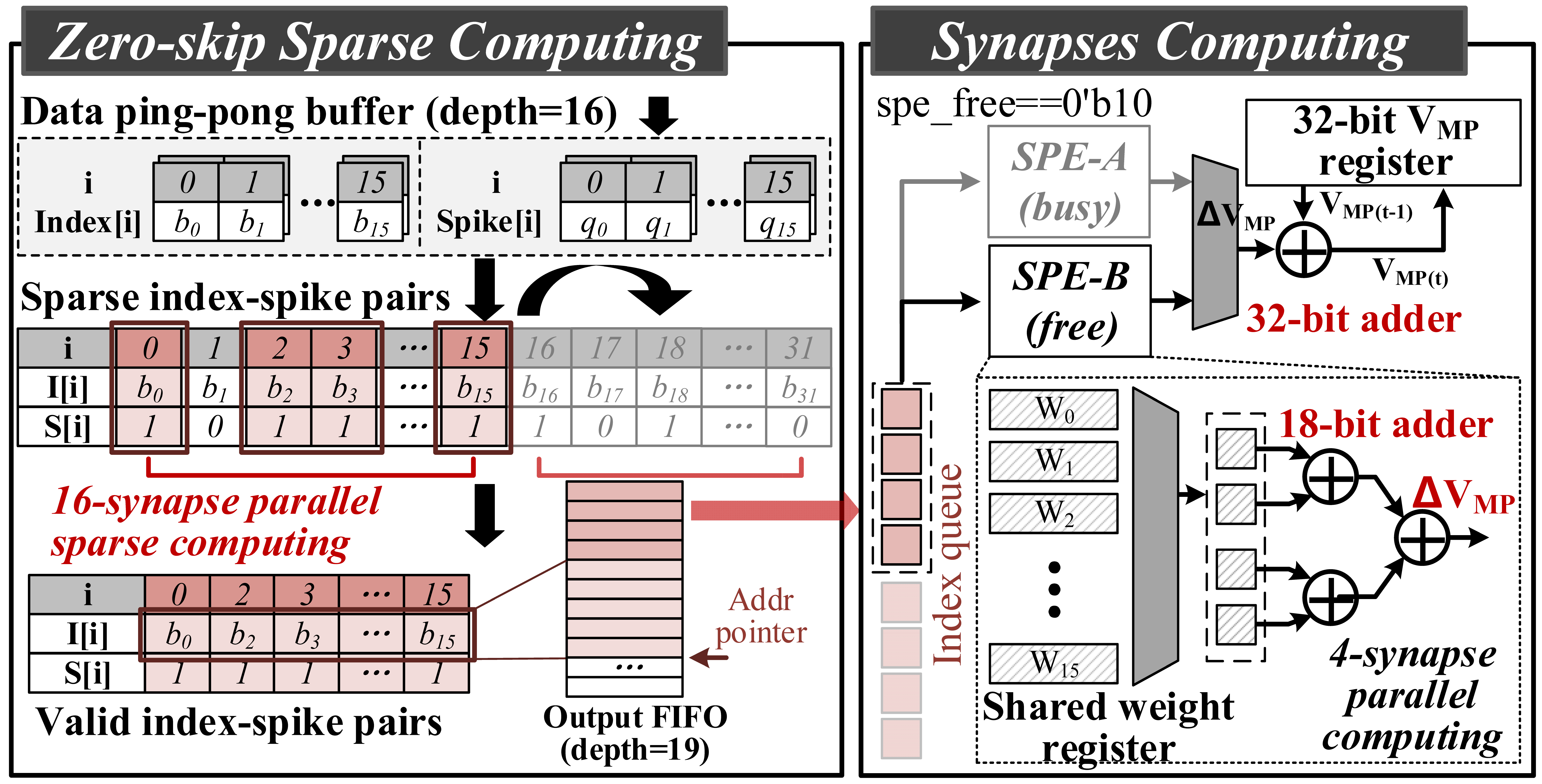}}
    \caption{Computing process of ZSPE and SPE.}
    \label{fig2}
\end{figure}
\begin{figure}[!t]
    \centerline{\includegraphics[width=\columnwidth]{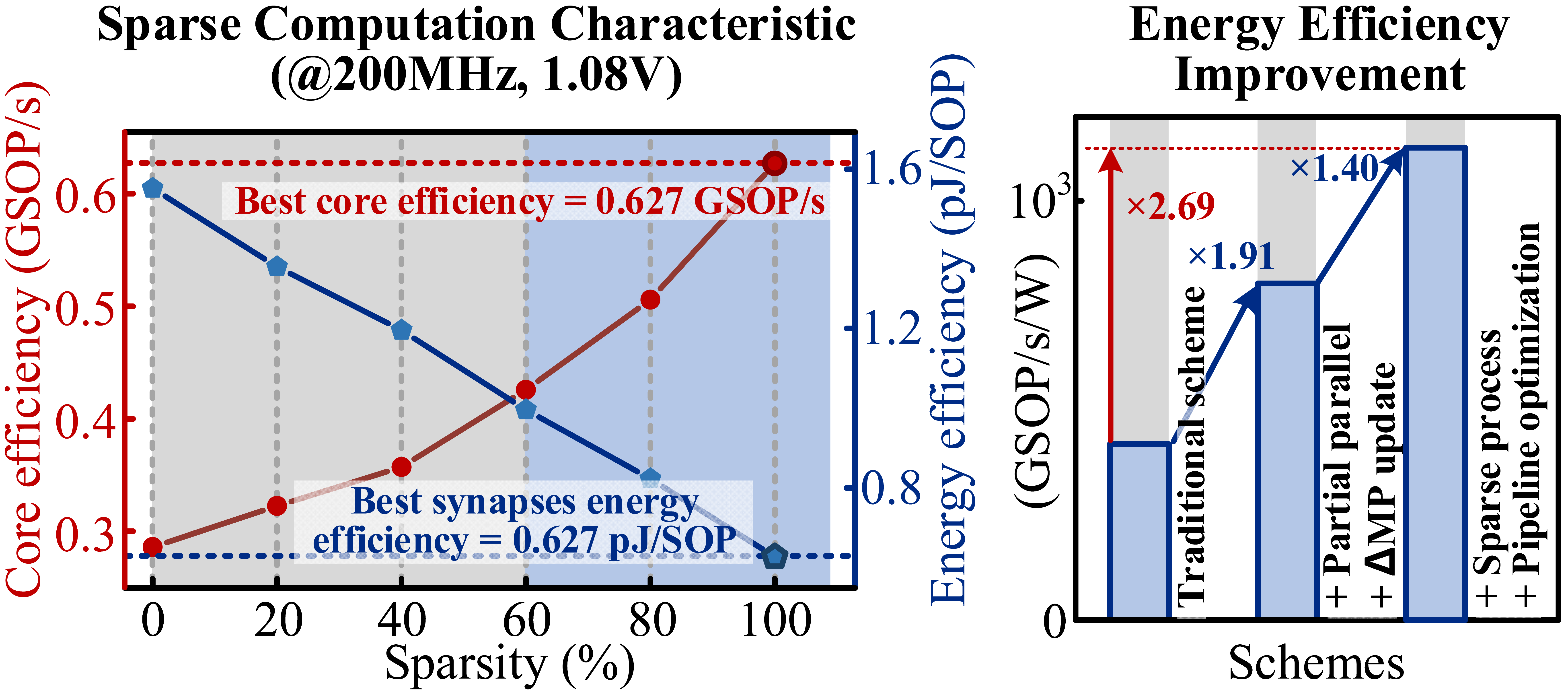}}
    \caption{Measurement of a neuromorphic core.}
    \label{fig4}
\end{figure}
Figure \ref{fig1} shows the neuromorphic core scheme. A clock gating enables the core clock according to an enable signal in the register table. In addition, the register table stores other parameters, such as neuron configuration parameters and read-only core ID. Double ping-pong caches facilitate expedited access to spike data and weight index. A zero-skip sparse process engine (ZSPE) enables 16-bit spike parallel sparse processing, and dual synapse process engines (SPE) realize 8-bit synapse computing (4-bit for each SPE). The computing process of ZSPE and SPE are shown in Fig. \ref{fig2}. All synapses share N$\times$W-bit quantized weights in a core, in which N is the weight number, and W is the weight bit width (\{N, W$\in$4,8,16\}). A neuron updater controls neuron MP integration, leaking and resetting, and spike firing procedures. 

A four-level pipeline is set up in the core, including core caches, ZSPE, SPE, and neuron updater. Buffers are inserted into the pipeline to optimize data-access efficiency. During computing, ZSPE loads 16 pre-spikes from caches, then initiates parallel sparse computing and forwards weight indexes of synapses with valid input spikes to SPE. SPE obtains four synapse weights according to the weight indexes and calculates the partial MP by parallel. Finally, the neuron updater accumulates the neuron MP then controls spike firing. 

Figure \ref{fig4} shows the measurement of the core design. The core computing efficiency and synapse energy efficiency are estimated over a spike sparsity range of 0-100\% under 200 MHz working frequency. The analysis indicates the neuromorphic core achieves the best computing and synapse energy efficiency of 0.627 GSOP/s and 0.627 pJ/SOP. Furthermore, the energy efficiency of the neuromorphic core is 2.69$\times$ of that of the baseline design with a traditional scheme.

\subsection{Fullerene-like NoC with Multi-mode Routers}
\begin{figure}[!t]
    \centerline{\includegraphics[width=\columnwidth]{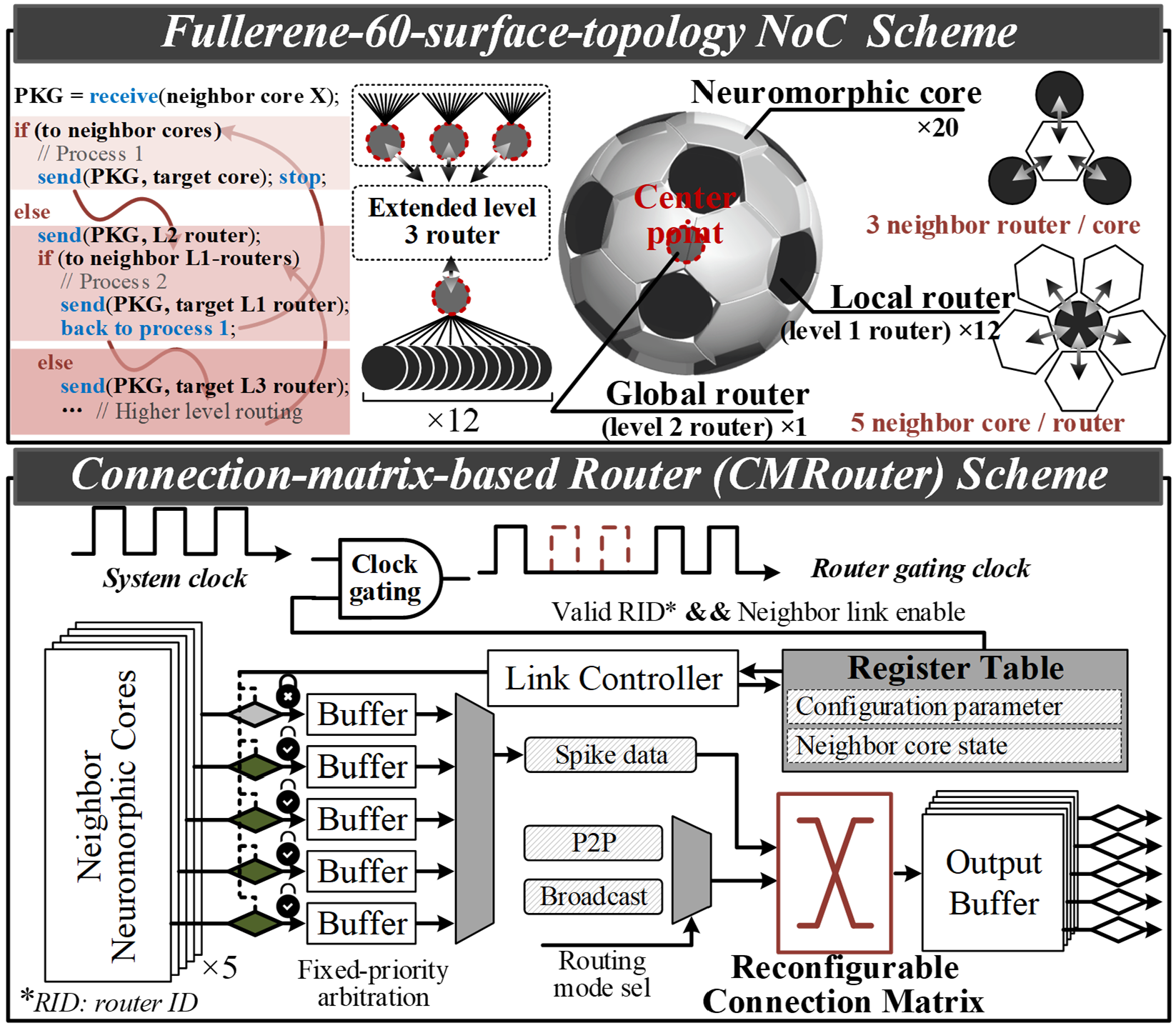}}
    \caption{Scheme of the fullerene-like NoC and CMRouter.}
    \label{fig5}
\end{figure}

Inspired by the fullerene-60, a fullerene-like NoC topology with multi-mode connection-matrix-based routers (CMRouters) is proposed, as shown in Fig. \ref{fig5}. Twenty cores and twelve level-1 routers form a level-1 fullerene-like routing domain. The center point of the topology is designed as the level-2 router for scaling up. The CMRouter (level-1 router) comprises independent input and output buffers, a register table, a link controller, a channel arbiter, a reconfigurable connection matrix, and a clock gating unit. The register table records the configured data-link parameters and neighbor core states. The link controller provides hang-up signals to input ports when data blocking or timestep is out of sync between cores. The connection matrix records all routing links among neighbor cores utilizing only $N_c\times N_c\times W_{cid}$ bit ($N_c$ is the neighbor core number of 5, and $W_{cid}$ is the width of the core id of 5) on-chip memory. The connection matrix allows the router to be compatible with multiple transmission modes, including P2P, broadcast, and merge, while avoiding complex packet encoding and decoding. 

\begin{figure}[!t]
    \centerline{\includegraphics[width=\columnwidth]{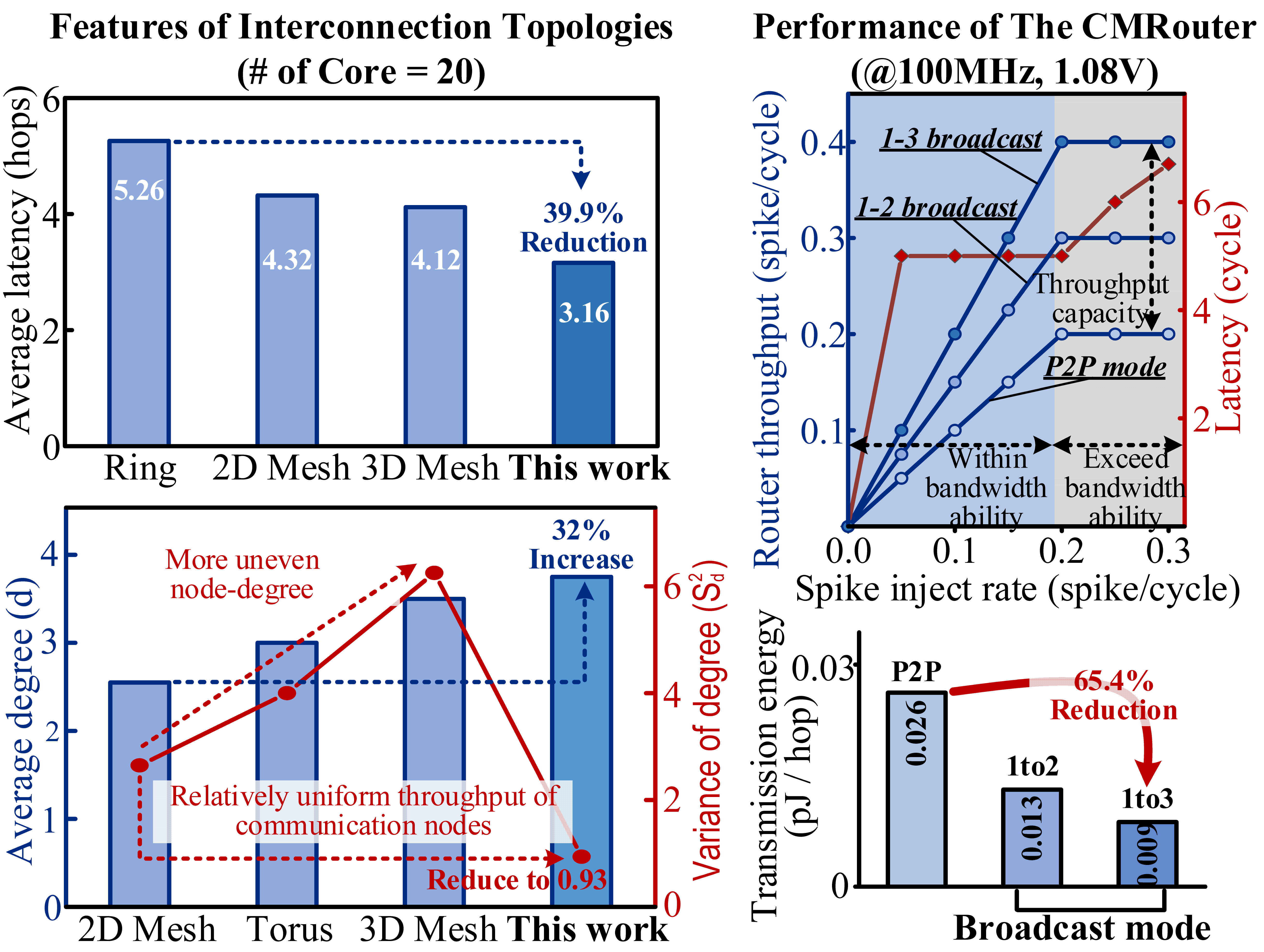}}
    \caption{Measurement of the NoC and the CMRouter.}
    \label{fig7}
\end{figure}

Figure \ref{fig7} shows the measurement of the NoC. The fullerene-like routing topology achieves a low average latency of 3.16 hops, better than other topologies (up to 39.9\%). In the fullerene-like topology, the average degree (d) of communication nodes reaches 3.75 (32\% better than 2D-mesh), allowing high-flexible data routing while effectively alleviating data transmission congestion. In addition, the variance of node degree ($S^2_d$) achieves 0.94, which is smaller than other topologies ($S^2_d\leq$ 2.6) and significantly guarantees more uniform throughputs of communication nodes in NoC. The CMRouter exhibits transmission energy efficiency of 0.026 pJ/hop and 0.009 pJ/hop in the P2P mode and 1to3 broadcast mode, respectively. Moreover, the CMRouter achieves a 0.2-0.4 spike/cycle throughput. 

\subsection{RISC-V CPU and Heterogeneous Integration}
\begin{figure}[!t]
    \centerline{\includegraphics[width=\columnwidth]{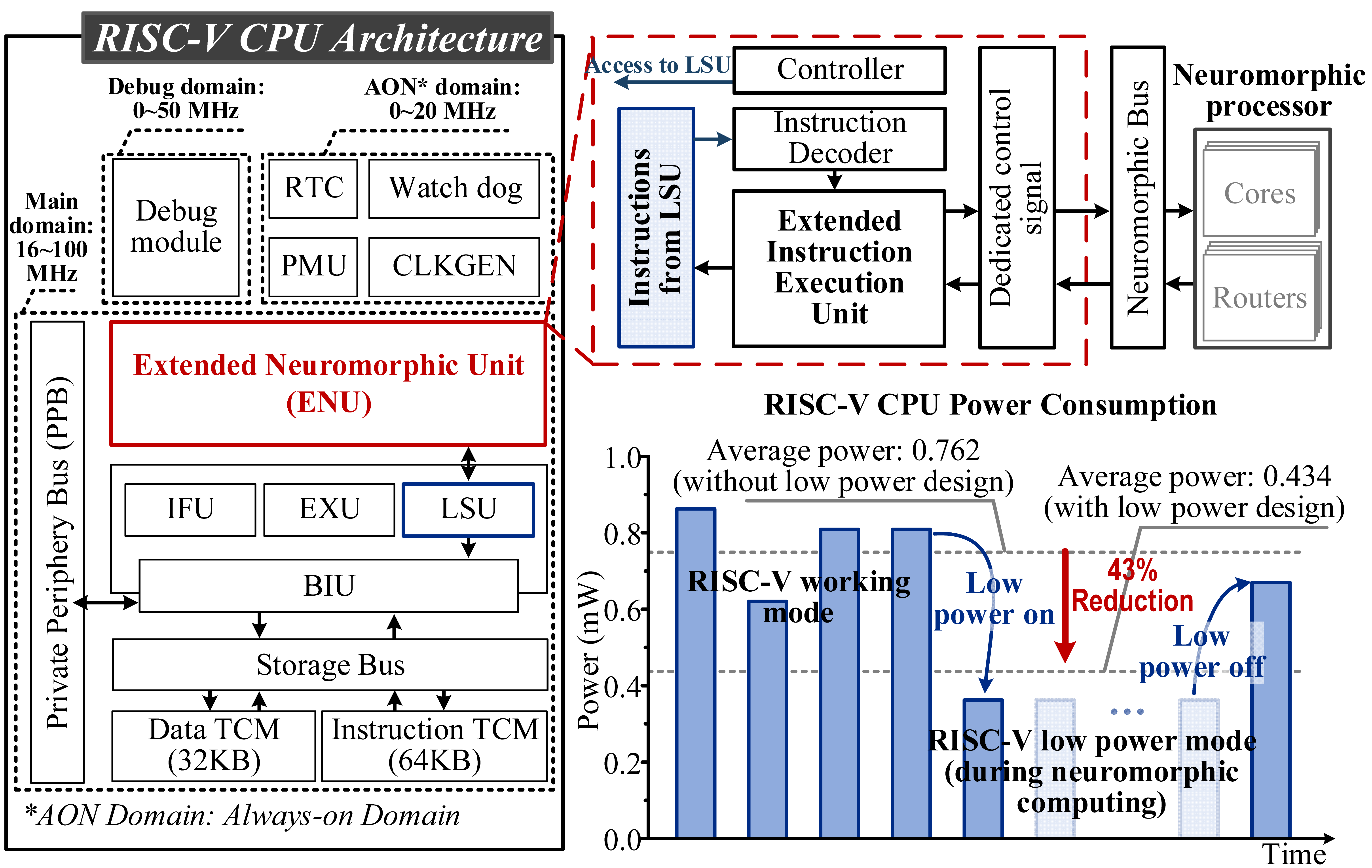}}
    \caption{Scheme of RISC-V CPU and the power measurement.}
    \label{fig8}
\end{figure}
To improve the flexibility of the chip and alleviate the dependence on peripheral circuits, a general-purpose processor (RISC-V CPU) and a neuromorphic processor are heterogeneously integrated together. The structure details of the RISC-V CPU and its power measurement are shown in Fig. \ref{fig8}. There are three different clock domains in the RISC-V core, in which the high-frequency clock (HFCLK) in the main domain can be halted by clock gating through a sleep instruction in software for low power. Meanwhile, the RISC-V core can be woken up through timestep-switch or network-computing-finish signals. The low power design enables RISC-V to achieve an average power consumption of 0.434 mW in the MNIST dataset, 43\% lower than the baseline.

An ENU is designed to couple the RISC-V CPU and neuromorphic processor, and a set of dedicated neuromorphic instructions (including network parameter initialization, core enable, network startup, etc.) has been extended for efficient control of the neuromorphic processor. For a tight coupling, the ENU and RISC-V core share a load-and-store unit (LSU) together. During working, the ENU controller sends an instruction access request to LSU, and then the LSU arbitrates the requests and sends the instruction to the ENU. The ENU generates dedicated control signals by decoding neuromorphic instructions and sends them to the neuromorphic processor through a neuromorphic bus.

\begin{figure}[!t]
    \centerline{\includegraphics[width=\columnwidth]{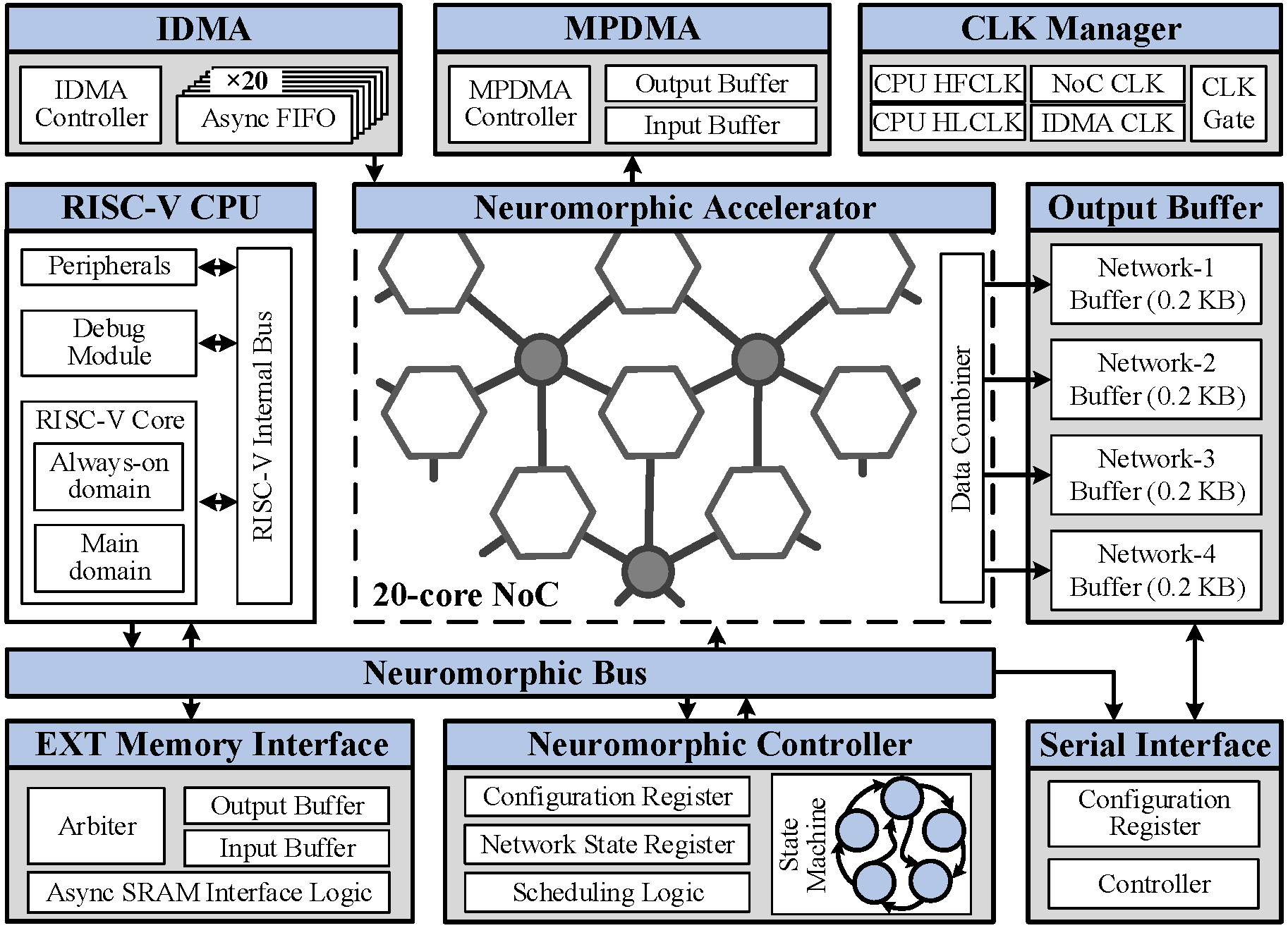}}
    \caption{Overall architecture of the SoC.}
    \label{fig10}
\end{figure}

\begin{table*}[htbp]
	\centering
    \caption{Comparison with SOTA neuromorphic chips for edge-AI application.}
    \renewcommand\arraystretch{1.4}
    \label{Table 1}
    \begin{tabular}{c|c|c|c|c|c|c}
    \toprule[2pt]
    \hline
            & ISSCC'23\cite{b9} & ISSCC'23\cite{b10} & ISSCC'22\cite{b11} & TBioCAS'22\cite{b5} & JSSC'20\cite{b4} & $\textbf{This work}$ \\
        \hline
        Technology & 28 & 28 & 28 & 55 & 28 & \textbf{55} \\
        \hline
        Cores & 64$\times$SNN & \makecell[c]{32$\times$CNN$+$\\32$\times$SNN} & 1$\times$SNN & 9$\times$SNN & \makecell[c]{156$\times$(ANN\&\\SNN)} & \makecell[c]{\textbf{\textcolor{blue}{1$\times$RISC-V+}}\\ \textbf{\textcolor{blue}{20$\times$ SNN}}} \\     
        \hline
        Application domain & \makecell[c]{Digital AI\\ computing} & \makecell[c]{Digital AI\\ computing} & \makecell[c]{Digital AI\\ computing} & \makecell[c]{Digital AI\\ computing} & \makecell[c]{Digital AI\\ computing} & \makecell[c]{$\textbf{\textcolor{blue}{General-purpose}}$\\ $\textbf{\textcolor{blue}{processing +}}$\\ $\textbf{\textcolor{blue}{Digital AI computing}}$}  \\
        \hline
        Supply voltage (V) & 0.56-0.90 & 0.7-1.1 & 0.5-0.8 & 1.2 & 0.85-1.15 & $\textbf{1.08-1.32}$ \\
        \hline
        Frequency (MHz) & 40-210 & 50-200 & 13-115 & 100 & 300 & \makecell[c]{$\textbf{16-100 (RISC-V)}$\\ $\textbf{50-200(AI processor)}$} \\
        \hline
        Die area (mm$^2$) & 1.63 & 20.25 & 0.86 & 6.00 & 14.44 & $\textbf{5.42}$ \\
        \hline
        Power (mW) & 2.91-56.8 & 32.3-294.1 & N. A. & 14.9 & 950 & $\textbf{\textcolor{blue}{2.8-113}}$ \\
        \hline
        \makecell[c]{Power density$^a$\\ (mW/mm$^2$)} & 1.79-34.85 & 1.6-14.52 & N. A. & 2.48 & 65.79 & $\textbf{\textcolor{blue}{0.52-20.85}}$ \\
        \hline
        \# Spiking neurons & 522 & 2 K & 272 & 9 K & 39 K & $\textbf{\textcolor{blue}{160 K}}$ \\
        \hline
        \makecell[c]{Neuron density$^b$\\ (\#/mm$^2$)} & 320.25 & 101.14 & 316.28 & 1.5 K & 2.8 K & $\textbf{\textcolor{blue}{30.23 K}}$ \\
        \hline
        \# Synapses & \makecell[c]{258 K\\ (8,10-bit)} & \makecell[c]{128 K\\ (4, 8-bit)} & \makecell[c]{132 K\\ (8-bit)} & \makecell[c]{9 M\\ (16-bit)} & \makecell[c]{9.75 M\\ (8-bit)} & \makecell[c]{$\textbf{\textcolor{blue}{1280 M}}$\\$\textbf{\textcolor{blue}{(4, 8, 16-bit)}}$} \\
        \hline
        \makecell[c]{Interconnection\\ topology} & Tree & N. A. & N. A. & Mesh & Mesh & $\textbf{\textcolor{blue}{Fullerene-like topology}}$ \\
        \hline
        Data routing mode & P2P & P2P & P2P & P2P & P2P & \makecell[c]{$\textbf{\textcolor{blue}{Hybrid}}$\\ $\textbf{\textcolor{blue}{(P2P, broadcast, merge)}}$} \\
        \hline
        Energy efficiency & 1.5 pJ/SOP & \makecell[c]{1.1 pJ/SOP\\ (ImageNet)} & \makecell[c]{5.3 pJ/SOP\\ (DVS Gesture)} & 33.3 pJ/SOP & 1.5 pJ/SOP & \makecell[c]{$\textbf{\textcolor{blue}{0.96 pJ/SOP (NMNIST)$^c$}}$\\ $\textbf{\textcolor{blue}{1.17 pJ/SOP (DVS Gesture)$^c$}}$\\ $\textbf{\textcolor{blue}{1.24pJ/SOP (Cifar-10)$^c$}}$} \\
        \hline
        Accuracy & \makecell[c]{96.0\% (NMNIST)\\ \makecell[c]{92.0\% (DVS\\Gesture)}} & \makecell[c]{77.1\%\\ (ImageNet)} & \makecell[c]{87.3\%\\ (DVS Gesture)} & \makecell[c]{99.5\% (NMNIST)\\ 97.2\% (DVS\\Gesture)} & \makecell[c]{98.22\%\\ (NMNIST)} & \makecell[c]{$\textbf{\textcolor{blue}{98.8\% (NMNIST)}}$\\ $\textbf{\textcolor{blue}{92.7\% (DVS Gesture)}}$\\ $\textbf{\textcolor{blue}{81.5\% (Cifar-10)}}$} \\
        \hline
    \end{tabular}
    \begin{tablenotes}
        \footnotesize
        \item[1] $^a$ Power density = chip power / chip area. $\qquad$ $^b$ Neuron density = \# of neurons / chip area. $\qquad$ $^c$ Working at 100 MHz, 1.08 V.
    \end{tablenotes}
\end{table*}

\subsection{Overall Architecture}
Figure \ref{fig10} presents the overall architecture of the chip integrating with the proposed neuromorphic processor and RISC-V CPU. The neuromorphic bus is used for interaction between the CPU and the neuromorphic processor. In addition, a neuromorphic controller and an external memory interface are also connected to the bus for processor control and off-chip asynchronous SRAM data access. Index DMA (IDMA) and membrane potential DMA (MPDMA) are used to transmit computing data to neuromorphic cores directly. A clock manager is responsible for the clock management of the system. Four independent 0.2 KB output buffers are used to store the computing results of different networks.

\section{Result and Comparison}
\begin{figure}[!t]
    \centerline{\includegraphics[width=\columnwidth]{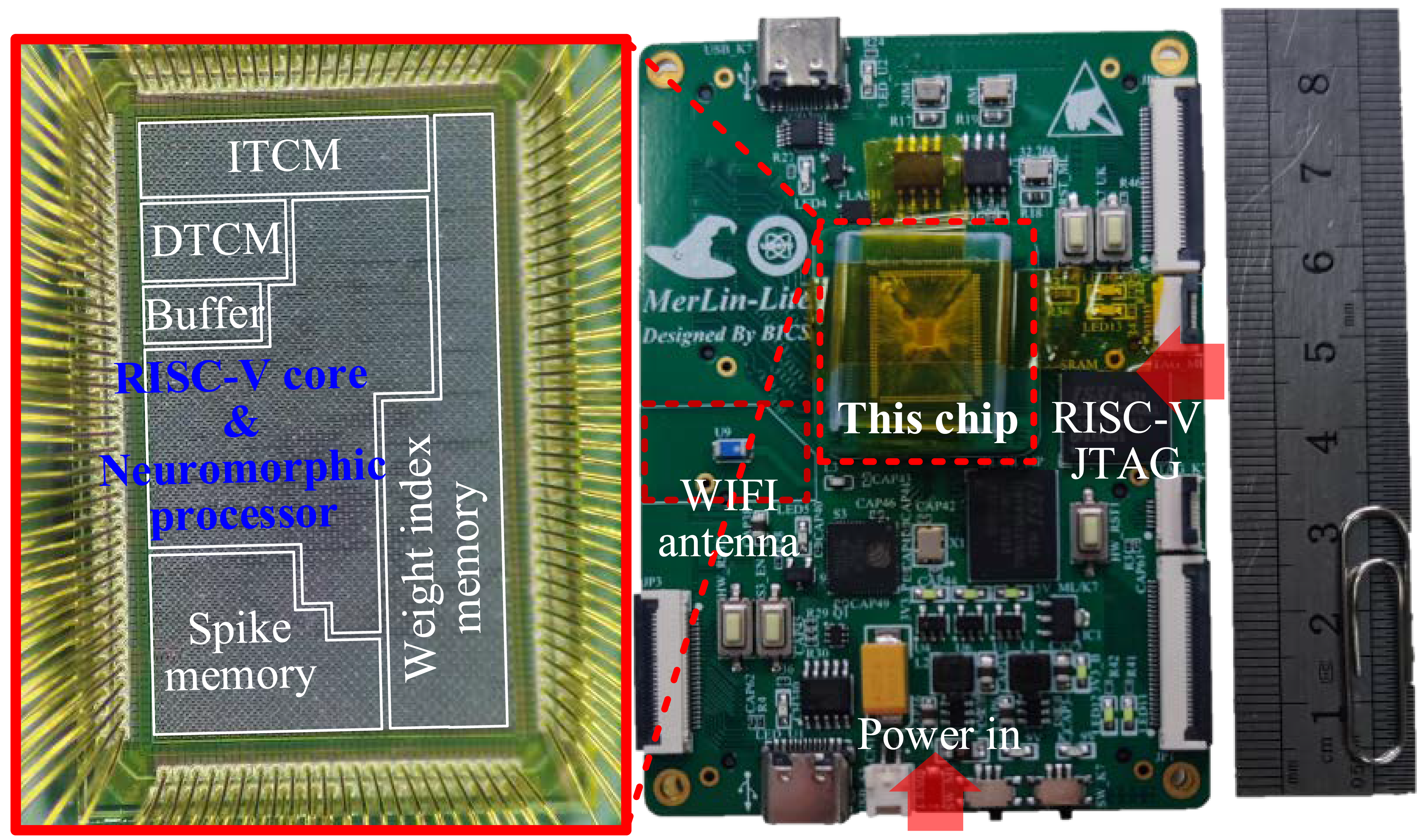}}
    \caption{Die photo and development platform.}
    \label{fig11}
\end{figure}

The neuromorphic chip is fabricated in a 55 nm CMOS process and occupies 5.42 mm$^2$ die area. Figure \ref{fig11} shows the die photo and development platform. The edge-AI platform allows high-energy-efficiency neuromorphic computing within an area of the human plam while enabling wired or wireless communication with other platforms. Based on the platform, the chip is demonstrated effective in different datasets, including NMNIST, DVS Gesture, and Cifar-10. Table \ref{Table 1} shows the comparison with related SOTA works. The chip consumes 2.8-113 mW at 1.08 V supply and has a minimum power density of 2.8 mW/mm$^2$. The neuromorphic core achieves a minimum of 0.96 pJ/SOP energy efficiency in applications. Furthermore, it has a high neuron density of 30.23 K/mm$^2$, 10$\times$ higher than previous works. 

\section{Conclusion}
This work proposed a 55 nm heterogenous neuromorphic SoC, including a RISC-V CPU and a neuromorphic processor for edge-AI computing, which has deeply optimized from core-, network-on-chip- and system-levels. The chip is demonstrated to be effective on different datasets and achieves a high energy efficiency of 0.96 pJ/SOP and a low power density of 0.52 mW/mm$^2$. It integrates a fullerene-like NoC with low average latency and high average node degree. It also has a high neuron density of  30.23 K/mm$^2$.
\section{Acknowledgement}
This work was supported by STI 2030-Major Projects 2022ZD0209700.

\newpage
\bibliographystyle{IEEEtran}
\bibliography{IEEEabrv, Neuromorphic}

\end{document}